\documentclass[10pt,aps,prb,superscriptaddress,twocolumn,showpacs,floatfix]{revtex4-1}
\usepackage{graphicx}
\usepackage{dcolumn}
\usepackage{bm}
\usepackage{amsmath}
\usepackage{subfigure}
\usepackage{url}
\usepackage{xfrac}

\begin{document}

\preprint{APS/123-QED}

\title{Coherent charge and magnetic ordering in Ho/Y superlattice revealed by element-selective x-ray scattering}

\author{V. Ukleev}
\email{victor.ukleev@psi.ch}
\affiliation{Laboratory for Neutron Scattering and Imaging (LNS), Paul Scherrer Institute (PSI), CH-5232 Villigen, Switzerland}
\author{V. Tarnavich}
\affiliation{National Research Center Kurchatov Institute -- Petersburg Nuclear Physics Institute, 188300 Gatchina, Russia}
\author{E. Tartakovskaya}
\affiliation{Institute of Magnetism National Academy of Sciences of Ukraine, Kiev 03142, Ukraine}
\affiliation{Kiev National University, Institute of High Technologies, Kiev 03022, Ukraine}
\author{D. Lott}
\affiliation{Institute for Material Research, Helmholtz-Zentrum Geesthacht, Max-Planck-Straße 1, 21502 Geesthacht, Germany}
\author{V. Kapaklis}
\affiliation{Department of Physics and Astronomy, Uppsala University, Box 516, SE-75120 Uppsala, Sweden}
\author{A. Oleshkevych}
\affiliation{Department of Physics and Astronomy, Uppsala University, Box 516, SE-75120 Uppsala, Sweden}
\affiliation{KTH Royal Institute of Technology, 100 44, Stockholm, Sweden}
\author{P. Gargiani}
\affiliation{CELLS Experiment Division, ALBA Synchrotron Light Source, Barcelona E-08290, Spain}
\author{M. Valvidares}
\affiliation{CELLS Experiment Division, ALBA Synchrotron Light Source, Barcelona E-08290, Spain}
\author{J. S. White}
\affiliation{Laboratory for Neutron Scattering and Imaging (LNS), Paul Scherrer Institute (PSI), CH-5232 Villigen, Switzerland}
\author{S. V. Grigoriev}
\affiliation{National Research Center Kurchatov Institute -- Petersburg Nuclear Physics Institute, 188300 Gatchina, Russia}
\affiliation{Saint Petersburg State University, 198504 St. Petersburg, Russia}
\date{\today}

\begin{abstract} 
Magnetic rare-earth / non-magnetic metal superlattices are well-known to display chiral spin helices in the rare-earth layers that propagate coherently across the non-magnetic layers. However, the underlying mechanism that preserves the magnetic phase and chirality coherence across the non-magnetic layers has remained elusive. In this Letter, we use resonant and element-specific x-ray scattering to evidence directly the formation of two fundamentally different long-range modulations in a Holmium/Yttrium (Ho/Y) multilayer: the known Ho chiral spin helix with periodicity 25\,\AA, and a newly observed charge density wave with periodicity 16\,\AA~that propagates through both the Ho and non-magnetic Y layer. With x-ray circular magnetic dichroism measurements ruling out the existence of a magnetic proximity effect induced moment in the non-magnetic Y layers, we propose that the charge density wave is also \emph{chiral}, thus providing the means for the transmittance of magnetic chirality coherence between Ho layers.
\end{abstract}


\maketitle

\section{Introduction}

Rare-earth (RE) metals are textbook examples of systems with indirect exchange coupling between localized $4f$ electron moments that is mediated by conduction electrons \cite{legvold1980rare,gschneidner2002handbook}. Due to their complex landscape of magnetic interactions, which includes Ruderman-Kittel-Kasuya-Yosida (RKKY) exchange, RE metals exhibit several intermediate magnetic textures between the high temperature ($T$) paramagnetic and lowest $T$ ferromagnetic phases \cite{legvold1980rare}, and are a rich playground for the study of chiral magnetism and magnetoelectric phenomena \cite{kishine2011tuning,togawa2012chiral,iwasaki2013universal}. In this context, the realization of nanoscale magnetic spirals in helimagnetic thin films and multilayers makes RE metals promising for use in high-density magnetic storage devices \cite{dor2013chiral,vedmedenko2014topologically,dzemiantsova2015stabilization}. Thus the deep understanding of the low energy instabilities in RE superlattices (SLs) provides unique insights into their rich behavior, in turn driving progress in magnetism based technology and spintronics.

In magnetic RE/non-magnetic Yttrium (Y) SLs, the coherent propagation of magnetic spin helices across the SL is well-known from neutron scattering \cite{rhyne1987occurrence,erwin1987magnetic,majkrzak1991magnetic}. Crucially, not only is the phase coherency of helical order preserved between RE layers, but the chirality is also preserved across the Y blocks \cite{salamon1986long,majkrzak1991magnetic,grigoriev2008field,tarnavich2014field}. In early works on RE/Y SLs \cite{rhyne1987occurrence,erwin1987magnetic} it was suggested that a helical spin density wave (SDW) is induced in the Y conduction electron band which couples via the RKKY interaction to local RE moments, thus preserving both the phase and chirality of magnetic spirals from layer to layer.

An alternative explanation was given by Yafet \emph{et al.}~\cite{yafet1988interlayer} suggesting that spiral coherence is maintained due to nesting features in the SL Fermi surface. For the case of helimagnetic RE/Y blocks, the intercalated Fermi surfaces of RE and Y layers, and the vector sum of the RKKY exchange fields induced by all atomic layers of the RE element, results in a non-trivial itinerant electron density distribution and a magnetization at the Y site. 
Thus, due to an anisotropic response of the Fermi surface nesting feature \cite{thakor2003first}, an elliptically polarized wave propagates through the Y medium, with a rotation sense determined by the RE magnetic spiral chirality. Although appealing, this theory does not convincingly explain certain experimental findings for Ho/Y, Ho/Lu, and Ho/Er SLs, such as the dependence of the transition $T$s on the thicknesses of the RE and the non-magnetic spacer layers, the $T$-independence of the spin coherence length, and the basal plane magnetic ordering \cite{simpson1994competing,cowley1994magnetic,cowley1998coherence}. Instead, a generalized theory of interlayer exchange coupling between magnetic layers through non-magnetic metallic spacer layers \cite{bruno1995theory} can explain the layer thickness dependence of the spin coherence \cite{cowley2004magnetic}, but its application to real systems with sophisticated Fermi surface topologies is highly non-trivial, particularly in the context of a propagating magnetic chirality. 

Alternatively, magnetic chirality can also be preserved across non-magnetic blocks in the presence of a noncentrosymmetric potential in the superlattice, owing, for example, to an asymmetric strain profile which favored one handedness. Such effects of strain- and defect-induced chirality are discussed in Refs.\cite{fedorov1997interaction,haraldsen2010control,yershov2015curvature,michels2018microstructural}. Indeed, we have shown recently that the presence of terraces on the surface of a sapphire substrate underneath a Ho/Y SL is responsible for inducing an effective uniaxial magnetic anisotropy, that is inhomogeneously distributed along the growth direction, resulting in a preferred handedness of the spin spirals \cite{tarnavich2017magnetic}. 

Understanding the mechanism of chirality preservation is thus of crucial importance in order to provide control of chiral magnetism in SLs and similar systems. Since neutron scattering did not detect the formation of a helical SDW in non-magnetic blocks of RE/Y SLs, a complementary investigation is highly desired. X-ray absorption spectroscopy (XAS) and resonant elastic x-ray scattering (REXS) are excellent element-selective tools that allow an unambiguous distinction between magnetic and electronic structures in RE and non-magnetic metal blocks. 

\begin{figure}
\includegraphics[width=8cm]{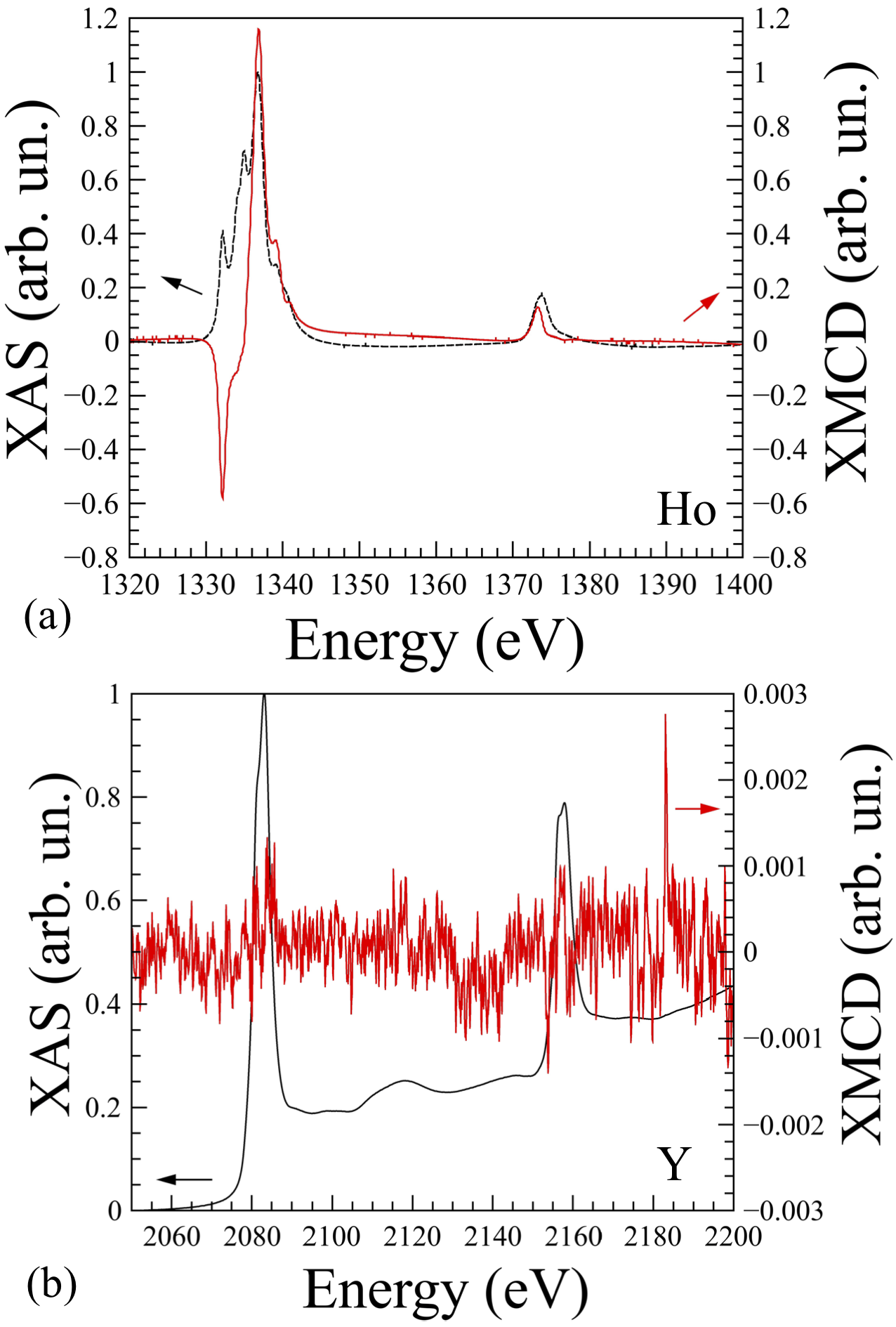}
\caption{XAS (dashed black line) and XMCD (solid red line) spectra measured near the absorption edges (a) $M_{4,5}$ of Ho and (b) $L_{2,3}$ of Y at $T = 30$\,K in an applied out-of-plane magnetic field of $H = 60$\,kOe.}
\label{fig1}
\end{figure}

\begin{figure*}
\includegraphics[width=16cm]{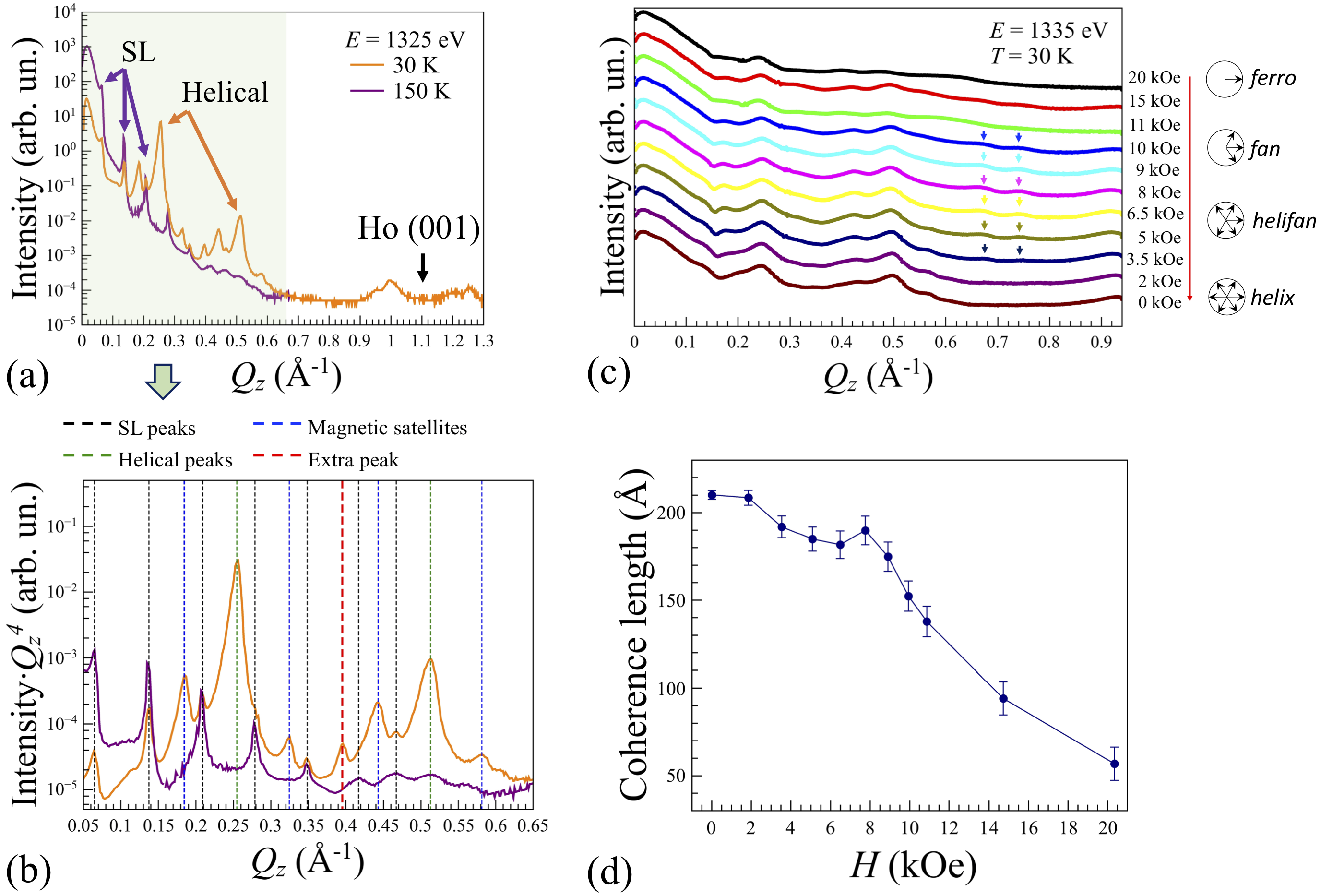}
\caption{(a) The REXS intensity as a function of momentum transfer vector $Q_z$ measured at the Ho $M_{5}$ pre-edge at 150\,K and 30\,K at zero field. (b) Identification of the charge and magnetic Bragg peaks. The REXS curves measured at $E=1325$\,eV are multiplied by $Q_z^4$ to compensate for the general decay of the reflectivity function. (c) Magnetic field evolution of the REXS curves at 30\,K corresponding to the transition from a fan to a helical state through an intermediate helifan phase. Higher-harmonic satellites around $Q_z=0.6-0.8$\,\AA$^{-1}$~are highlighted by arrows as markers of helifan and fan phases. Schematic basal-plane magnetic moment distributions for different magnetic structures are given in the right panel. (d) Coherence length of the magnetic modulation as a function of applied in-plane magnetic field.}
\label{fig2}
\end{figure*}

\section{Experimental}

Here we report XAS, x-ray magnetic circular dichroism (XMCD) and REXS experiments aimed at identifying long-period density wave modulations in a Ho/Y SL consisting of 30 periods of [Ho 60\,\AA/Y 30\,\AA]. The SL was grown by molecular beam epitaxy on a sapphire substrate with a Nb buffer layer. Further details of the sample preparation and characterization are reported elsewhere \cite{tarnavich2014field,tarnavich2014structural}.

XAS and XMCD experiments were carried out at the HECTOR endstation of the BOREAS beamline at the ALBA synchrotron (Barcelona, Spain). The beamline is dedicated to polarization-dependent x-ray spectroscopy in a wide energy range from 80 to 4500\,eV, that covers the $M_{4,5}$ edges of Ho and the $L_{2,3}$ edges of Y, respectively \cite{barla2016design}. Details of the measurement protocols are given in the Appendix A. To collect the XMCD signal, the samples were cooled to $T=30$\,K, well below the ordering temperature $T_N=118$\,K and a magnetic field $H$ of 60\,kOe was applied perpendicular to the film plane. This condition corresponds to the field-induced ferromagnetic phase of the SL. Figure \ref{fig1}a shows the averaged XAS and XMCD spectra measured at the Ho $M_{4,5}$ edges, with XAS spectra measured with opposite polarizations shown in see Appendix A.
The XAS line shape is characteristic to both the $3d \rightarrow 4f$ transitions (dashed line in Fig. \ref{fig1}a) expected for the $4f^{10}$ occupation of Ho metal \cite{thole19853d}. The observation of a large XMCD (solid red curve in Fig. \ref{fig1}a) clearly indicates the field-induced ferromagnetic state of Ho.

Figure \ref{fig1}b shows the XAS spectrum measured across Y $L_{2,3}$ edge that lies between $E=2050$ and $2200$\,eV, and which is characteristic to $2p \rightarrow 4d$ transitions. The XMCD spectrum was also measured to probe a possible spin polarization of the Y $4d$ shells. In contrast to the RE element, no significant difference between the absorption spectra for right- and left-circularly polarized x-rays was observed (solid red curve in Fig. \ref{fig1}b). The estimated upper limit of the XMCD/XAS intensity ratio is tiny at 0.1\%. Given that much stronger XMCD signals were observed previously at the Y $L_3$-edge in Y$_3$Fe$_5$O$_{12}$ (YIG) films (XMCD/XAS\,$ \approx 1.3\%$) \cite{rogalev2009element} and DyFe$_2$/YFe$_2$ multilayers (XMCD/XAS\,$\approx 2.4\%$) \cite{dumesnil2009temperature} even at room temperature, we conclude that the Y blocks in the [Ho 60\,\AA/Y\,30 \AA] SL do not possess any significant spin polarization. Using the XMCD signal observed from Y in the YIG system, which was estimated to correspond to $\mu_Y \approx0.03$\,$\mu_B$, as a proportional rule, the noise in the signal observed at the Y $L_{2,3}$ absorption edges in the present system limits the possible moment induced in the Y $d$-band at just $\mu_Y<0.0025$\,$\mu_B$.

\begin{figure*}
\includegraphics[width=16cm]{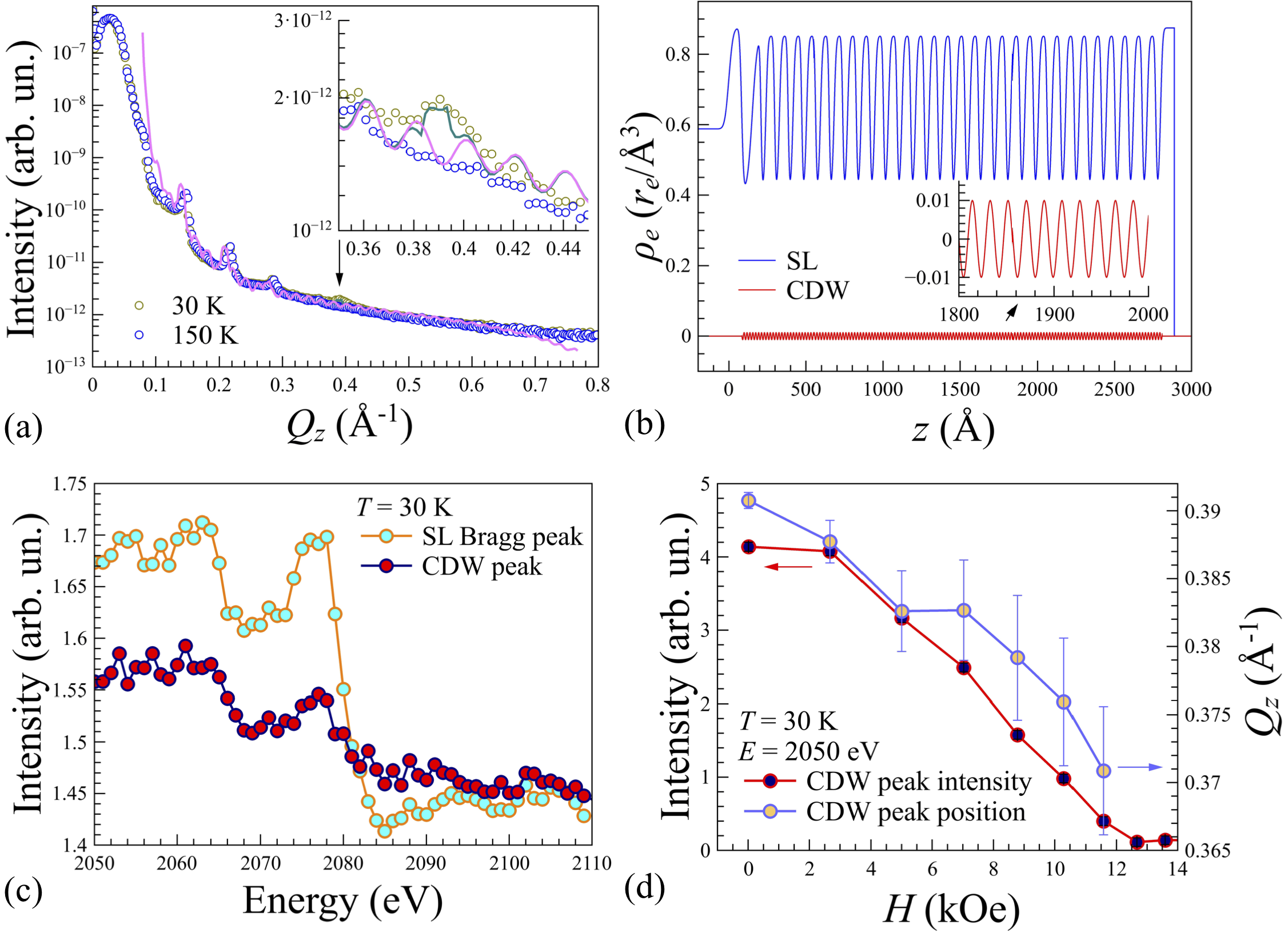}
\caption{(a) REXS curves measured at $E=2050$\,eV (Y $L_{3}$ pre-edge) above and below the magnetic ordering temperature: $T=150$\,K and $T=30$\,K at zero field. The simulated REXS curves calculated from Fourier transforms of $\rho(z)$ for the SL without (light magenta line) and with (dark cyan line) an additional CDW modulation. The corresponding electron density profiles are shown in the panel (b). (c) Energy dependence of the REXS intensity of the 4$^{th}$ order superlattice (SL) Bragg peak and incommensurate CDW peak near $Q_z\approx 0.39$\,\AA$^{-1}$~across Y $L_{3}$ edge. (d) Dependence of the intensity and position of the CDW peak with applied $H$.}
\label{fig3}
\end{figure*}

The REXS measurements were carried out using the MaReS endstation at the BOREAS beamline. Further details of the setup are given in the Appendix B. Above the ordering $T$ the SL structure was studied by measuring the specular reflectivity curve as a function of the momentum transfer vector $Q_z$. Figures~\ref{fig2}a,b show that five orders of SL structure peaks are observed at $nQ_z=n2\pi/ d_{SL}$, where $d_{SL}$ is the SL period and $n=1,2,3,4,5$, thus demonstrating the good quality of the SL structural order. Below $T_N$ at 30\,K, strong small-angle magnetic scattering and magnetic satellites appear near the forbidden Ho (001) Bragg peak (orange curves in Figures \ref{fig2}a,b). An identification of features in the REXS profile measured at the pre-edge energy $E=1325$\,eV is also given in Figure \ref{fig2}a, and is consistent with neutron data obtained from the same sample \cite{tarnavich2014structural}. Furthermore, energy dependent REXS data reproduce the findings of previous soft x-ray scattering results obtained on Ho films \cite{spencer2005soft,ott2006magnetic} (see Appendix B for more details). 

To obtain more insight into the features in the scattering patterns, in Figure \ref{fig2}b we show the REXS data from Figure \ref{fig2}(a) multiplied by $Q_z^4$ to compensate for Fresnel decay. The identified charge and magnetic Bragg peaks are in accord with previous neutron scattering data \cite{tarnavich2014structural}. Besides the SL Bragg peaks and the main magnetic peaks originating from the Ho helix (black and green dashed vertical lines, respectively) pairs of magnetic satellites are observed symmetrically on the left and right sides of each main magnetic peak due to the SL periodicity. The periodicity of the magnetic structure in the Ho layers at zero $H$ is found to be $\lambda_{Ho}=25$\,\AA~with a coherence length $\xi$ $\sim$ 510\,\AA. The value of $\xi$ corresponds to about 5\,periods of the SL, and is in excellent agreement with the SL structural coherence length of about 500\,\AA~ reported previously \cite{tarnavich2017magnetic}. An additional peak that cannot be explained by the helical or structural order of SL appears at $Q_z = 0.39$\,\AA$^{-1}$~below $T_N$. 
The nature of the new peak observed at $Q_z = 0.39$\,\AA$^{-1}$ will be further discussed in the context of REXS results measured at Y $L_3$ edge.

The $H$-dependence of REXS curves at the Ho $M_5$ edge was collected upon $H$ decreasing from 20\,kOe to 0\,kOe (Fig.\ref{fig2}c) at $E=1335$\,eV. In this field range a continuous transition from a fan to a helical phase via an intermediate helifan phase takes place, as observed by neutron diffraction \cite{de1999magnetic}. In the present experiment, the appearance of higher-harmonic satellites around $Q_z=0.6-0.8$\,\AA$^{-1}$~ at fields of $H\approx 3.5-10$\,kOe indicates the magnetic structure to deviate from a helix towards a helifan, this being fully consistent with the previous results \cite{jehan1992magnetic,de1999magnetic}. Interestingly, the magnetic coherence length $\xi$ of the modulation tends to decrease as a function of $H$ (Fig. \ref{fig2}d). When precisely at the resonant energy the finite penetration depth of the soft x-rays leads to a broadening of the Bragg peaks, thus the observed $\xi$ deviates from the value of $510$\,\AA~measured just off-resonance at $E=1325$\,eV. Nevertheless, the gradual shortening of magnetic coherence is clear for fields $H>8$\,kOe. De la Fuente \emph{et al.} suggested that $\xi$ for the helifan and fan phases can be shorter than for the pure helix in zero $H$ due to an enhanced scattering of conduction electrons from uncompensated ferromagnetic moments \cite{de1999magnetic}. This mechanism is similar as for antiferromagnetically coupled bilayers in giant magnetoresistive spin valves \cite{bruno1995theory}.

To directly probe spin correlations in the conduction band of non-magnetic Y layers, we performed REXS experiments at the Y $L_3$ edge. The scattering intensities collected at the pre-edge energy $E=2050$\,eV above ($T=150$\,K) and below ($T=30$\,K) $T_{N}$ are shown in Fig.\ref{fig3}a. After cooling from 150\,K to 30\,K an additional peak appears at $Q_z=0.39$\,\AA$^{-1}$~corresponding to a real-space modulation with a periodicity of $\lambda_Y=16$\,\AA. Besides the SL structural Bragg peaks, the peak at $Q_z=0.39$\,\AA$^{-1}$~is the only common feature observed in both REXS patterns measured at the Ho and Y edges below $T_N$.

\section{Results and discussion}

Considering the estimated upper value of the Y magnetic moment of $\mu_Y<0.0025$\,$\mu_B$ we rule out a magnetic contribution to the REXS pattern in Figure~\ref{fig3}(a) since it would be too small compared to the Thompson scattering from the electron density modulation in the SL. Since no XMCD is observed for Y, the optical theorem and Kramers-Kronig relations imply that both imaginary and real parts of the refractive index have vanishingly small magnetic contributions \cite{fink2013resonant}. Thus instead of any incommensurate magnetic modulation, our observations suggest strongly that the peak at $Q_z=0.39$\,\AA$^{-1}$ originates from Thomson scattering on itinerant electrons or the associated lattice distortions, i.e. a charge-density-wave (CDW). This CDW peak was not observed previously by neutron reflectometry done on the same sample \cite{tarnavich2014field}, which indicates the neutron scattering cross-section of the CDW-driven periodic distortion to lie below the measurement sensitivity. In REXS measurements, the onset of a CDW peak is often accompanied by Thompson scattering due to the associated long-period structural distortion of the lattice \cite{gibbs1985magnetic,abbamonte2006charge} and can be confirmed by means of polarization analysis \cite{gibbs1985magnetic}. 

To explore the possible CDW in more detail, we fitted the REXS data measured above $T_{N}$ and obtain the structural electron density profile of the SL (see Appendix C). To describe the additional peak at $T=30$\,K we simulated the REXS curves expected by calculating the Fourier transforms of the electron density profiles $\rho(z)$ corresponding to when an additional harmonic CDW is coherently propagating through the SL, and also when it is absent \cite{bataillou2003direct}. The resulting curves, multiplied by Fresnel decay $Q_z^{-4}$, are shown in Figure \ref{fig3}a. The $\rho (z)$ profiles of the SL and CDW used in the model are shown in Figure \ref{fig3}b. The profile due to the CDW with periodicity of 16\,\AA~and magnitude of 0.01\,$r_e$/$\AA^3$, where $r_e$ is the classical electron radius, combined with the SL profile provides a semi-quantitative description of the entire REXS curve. The amplitude of the simulated CDW is furthermore reasonable assuming the estimated density of conduction electrons $\sim 0.014$\,$r_e$/\AA$^3$ for Y metal \cite{reale1973concentration}.

Finally, to confirm the charge nature of the modulation in the Y layers we performed an energy scan around the peak near $Q_z=0.39$\,\AA$^{-1}$~(Fig. \ref{fig3}c). Overall therefore, we conclude that in Ho/Y SLs the magnetic phase coherence between neighboring Ho layers is preserved by a CDW induced in the conduction band of the metallic Y and not by a helical SDW as proposed in Refs. \cite{rhyne1987occurrence,erwin1987magnetic}. The CDW coherence length across the SL is $\sim$400\,\AA, being of similar order with the magnetic coherence length of the helical modulation in the RE blocks at $H$=0.

The intensity and position of the CDW peak are shown as a function of $H$ in Figure \ref{fig3}d. The gradual decrease of the intensity clearly indicates a field-induced suppression of the CDW, that is finally achieved at 12\,kOe. This field corresponds to the helifan-fan magnetic phase transition in Ho, and matches the field region where coherency is lost in the SL. The data further suggest that long-range interactions in SLs are only maintained in RE/Y superlattices if both the RE and spacer metals have similar or compatible band structures, and the symmetry of the spin ordering is compatiable with an initial itinerant CDW. Therefore, the Ho helifan and fan phases, both being $H$-induced magnetic structures, do not propagate across the Y layers. An appropriate ab-initio calculation of the band structure of magnetic SLs including several atomic layers of RE and Y is highly desired to confirm the existence of the required twisting of electron density in the Y conduction band.

Interestingly, the turn angle between adjacent atomic planes in the Y layers with a corresponding wave vector of 0.31\,\AA$^{-1}$~is found from a theoretical description of neutron diffraction experiments to be typical for many dilute RE alloys and RE/Y superlattices \cite{erwin1987magnetic,bohr1989diffraction,jehan1993magnetic}. In our direct measurement the wave vector of the modulation in the Y layers is 0.39\,\AA$^{-1}$~with a correspondingly shorter periodicity and a turn turn angle of 64$^\circ$. Our experimental observation is close to the value of a calculated susceptibility peak for Y metal at (0,0,0.375)$2\pi/c=0.41$\,\AA$^{-1}$~predicted by Liu et al. \cite{liu1971generalized}. On the other hand, more recent positron annihilation experiments \cite{dugdale1999nesting} and theoretical calculations give the position of a nesting feature at the Fermi surface to be at (0,0,0.55$\pm$0.02)$\pi/c=0.30$\,\AA$^{-1}$~for Y and diluted Y-RE alloys \cite{thakor2003first}. We suspect that the discrepancy between the periodicities previously proposed nesting features and the observed CDW periodicity might originate from a different Fermi surface shape in Ho/Y SL compared to bulk Y metal. Indeed, the CDW peak is observed at the Ho $M_5$ edge too (Fig. \ref{fig2}b), showing a coherent propagation of the itinerant wave across the RE and Y layers. 

\begin{figure*}
\includegraphics[width=16cm]{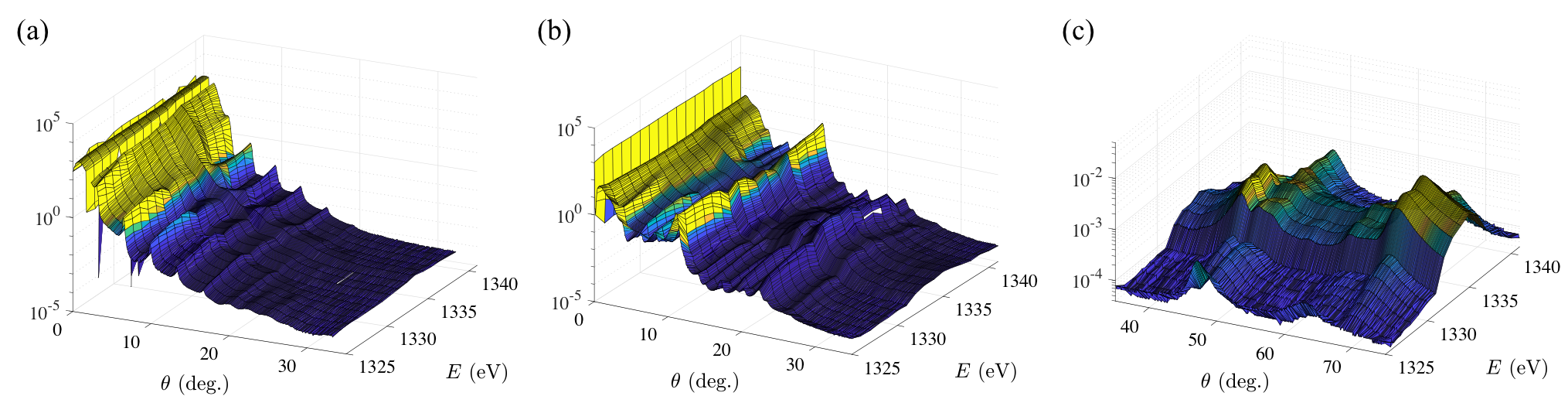}
\caption{(Color online) Energy dependence of the resonant elastic x-ray scattering (REXS) intensity measured across the Ho $M_{5}$ at zero field at 300\,K (a) and 30\,K (b). (c) Magnification of the REXS energy dependence near Ho (001) Bragg peak measured at $T = 30$\,K. Magnetic satellites are strongly enhanced near the resonant condition.}
\label{figS2}
\end{figure*}

\section{Conclusion}

To summarize, from element-selective resonant x-ray spectroscopy and scattering we identify two different types of modulation in a Ho/Y superlattice: a magnetic helix in the Ho layers that is accompanied by a second modulation due to a charge density wave (CDW) in non-magnetic Y layers. We assume that non-trivial $q$-dependent electron-phonon coupling plays an indispensable part in formation of CDW in both Y and Ho layers. As it has been shown in the pioneering resonant x-ray scattering investigations, magnetic ordering in rare-earth metal induces incommensurate elastic distortions \cite{gibbs1985magnetic}, which is indistinguishable to a CDW \cite{johannes2008fermi}. Since a scalar CDW cannot maintain any information of the handedness of a neighboring chiral density wave, we suggest that the coherence and chirality of the magnetic modulation between Ho layers is preserved via a \textit{chiral} itinerant CDW induced in the Y conduction electron band that couples via a non-trivial RKKY interaction to the Ho magnetic moments. Indeed, the multipolar electric charge ordering has been observed in bulk Ho \cite{yokaichiya2004observation}. Propagation of the multipolar charge order into Yttrium layers might be responsible for the chiral charge modulation. 

Thus, we rule out a magnetic proximity effect accompanied by a spin density wave in Y layers as a means for the transmittance of the magnetic chirality as proposed theoretically long ago \cite{rhyne1987occurrence,erwin1987magnetic}. While we cannot exclude the possibility that magnetic chirality may also be maintained between rare-earth layers by another mechanism, for example via magneto-elastic coupling due to the asymmetric strain profile, or defect-induced Dzyaloshinskii-Moriya interactions, the present results reinvigorate the debate, and provide new qualitative and quantitative aspects for the understanding of the RKKY interactions in rare-earth superlattices that can motivate further theoretical and experimental developments.

\section*{Acknowledgements}

We thank ALBA synchrotron radiation facility for the opportunity to carry out the soft x-ray scattering experiments as a part of the Proposals No. 2017092410 and 2019033575. V.U. and J.S.W. acknowledge financial support from the SNF Sinergia CRSII5\_171003 NanoSkyrmionics and SNF 200021\_188707. V.Tarnavich acknowledges financial support from the RFBR 18-32-00005 mol\_a, which allowed the preparation and certifying of the sample. V.K. would like to acknowledge the Knut and Alice Wallenberg foundation for financial support under Project No. 2015.0060, the Swedish Research Council and the Swedish Foundation for International Cooperation in Research and Higher Education. M.V. acknowledges additional funding for the scattering endstation by Grants MICINN ICTS-2009-02 and by MINECO FIS2016- 78591-C3-2-R (AEI/FEDER, UE).

\section*{Appendix A: X-ray absorption spectroscopy}

\begin{figure}
\includegraphics[width=8cm]{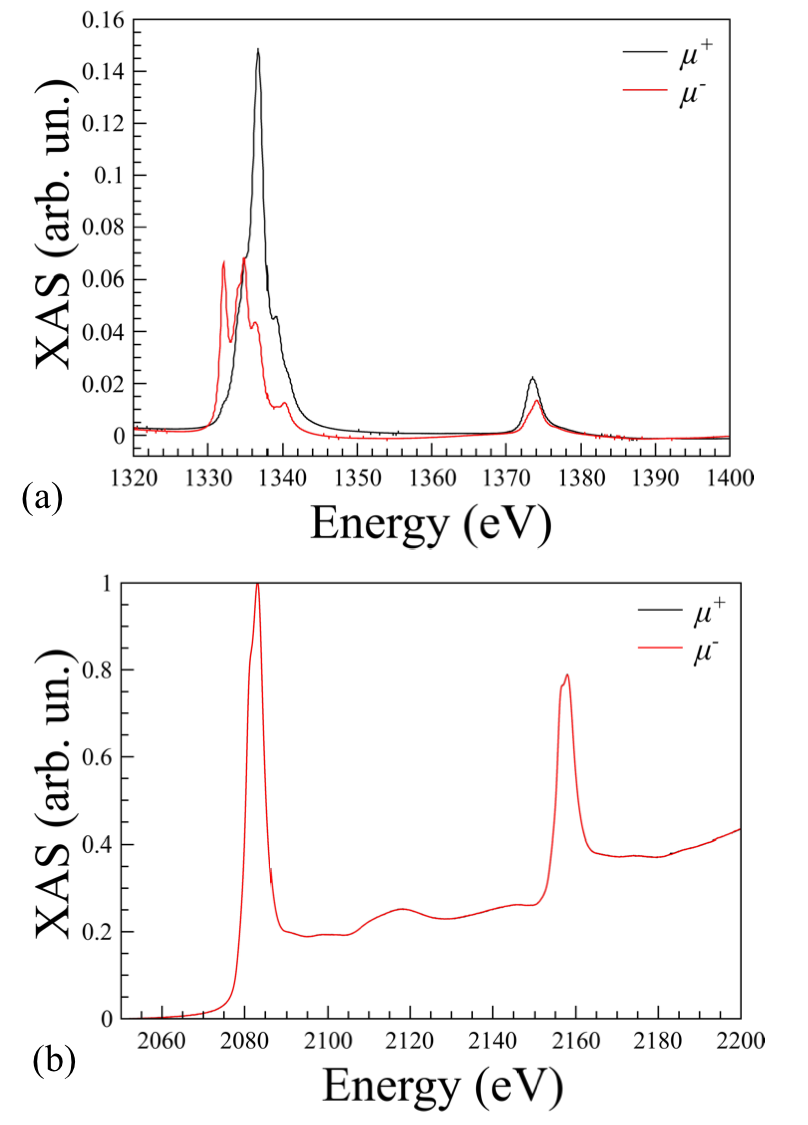}
\caption{(Color online) XAS and XMCD spectra at Ho $M_{4,5}$ (a) and Y $L_{2,3}$ (b) edges at $T = 30$\,K in an applied out-of-plane magnetic field of $H = 60$\,kOe measured with opposite $\mu^+$ and $\mu^-$ polarizations.}
\label{figS1}
\end{figure}

X-ray absorption (XAS) and x-ray magnetic circular dichroism (XMCD) measurements were carried out by detecting XAS spectra at the Ho $M_{4,5}$ and Y $L_{2,3}$ edges using left and right circularly polarized light in the total electron yield (TEY) mode. At Ho and Y the spectra were measured with 90\% circular polarization using third undulator harmonic. XAS intensity was normalized to the incoming photon flux measured via the drain current of a gold mesh monitor on the beam path upstream the sample. The temperature and magnetic field were controlled by a vector cryomagnet. The XMCD signal was calculated as the difference between the XAS spectra recorded for parallel ($\mu^+$) and antiparallel ($\mu^-$) polarizations of the photon helicity with the applied field $H$ (Fig. \ref{figS1}). Fields of $\pm 6$\,T were used to magnetize the sample perpendicularly to the film plane. The obtained pair of absorption spectra for parallel and antiparallel orientation of photon spin and sample magnetization determines the averaged XAS and XMCD signals, given by XAS = 1/2($\mu^+$ + $\mu^-$) and XMCD = $\mu^+$ - $\mu^-$, respectively.

\section*{Appendix B: X-ray scattering at the absorption edge of Holmium}

The REXS was measured near Ho $M_{5}$ edge above and below the ordering temperature (Fig.\ref{figS2}). At the absorption maximum ($E=1330-1340$\,eV) the charge scattering is strongly suppressed and magnetic scattering is enhanced. Due to the strong absorption at the resonant condition the penetration depth of the soft x-ray beam was limited by a few periods of superlattice what explains broadening and intensity decrease of the small-angle magnetic peaks. Thus, at the angle of incidence at the 1$^{st}$ order magnetic peak position and energy 1335\,eV a penetration length is only of a few hundreds \AA \cite{spencer2005soft}. Strong enhancement of magnetic satellites in resonant condition $E=1330-1340$\,eV reproduces previous soft x-ray scattering observations for Ho film (Fig. \ref{figS2}) \cite{spencer2005soft,ott2006magnetic}. 

\section*{Appendix C: Structural profile of the superlattice}

\begin{figure}
\includegraphics[width=8cm]{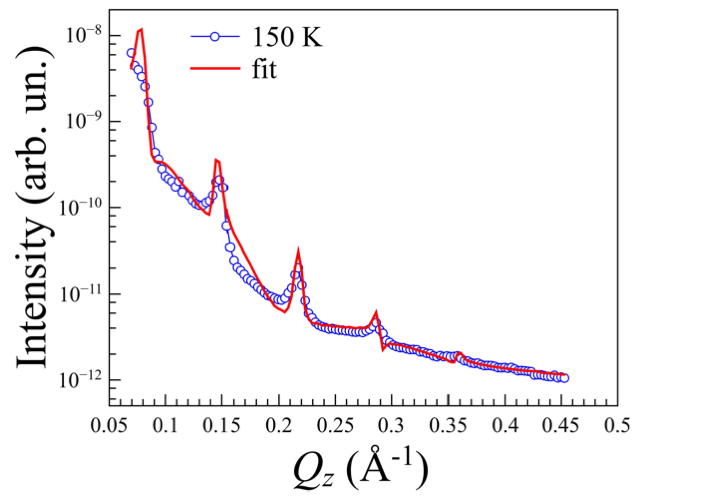}
\caption{(Color online) REXS curve measured at $E=2050$\,eV above the magnetic ordering temperature ($T=150$\,K) zero field (symbols) and fitted model (line) corresponding to the electron density profile shown in Fig. 3 of the main text.}
\label{figS3}
\end{figure}

To obtain the structural electron density profile $\rho (z)$ of the superlattice we have fitted the REXS data measured above the magnetic ordering temperature $T=150$\,K (Fig. \ref{figS3}a) by using GenX software \cite{bjorck2007genx}. Further, we have calculated the REXS curve for the SL containing additional harmonic electron density wave (or charge density wave, CDW) coherently propagated through the structure by calculating Fourier transform of the electron density profile (Fig. \ref{figS3}b) multiplied by Fresnel decay function $Q_z^{-4}$. Resultant electron density profiles of the SL and CDW are shown in Fig. \ref{figS3}c.

\end{document}